# A New Basis for Interpretation of the Planck Length

C. L. Herzenberg

**Abstract**
*A critical length has recently been identified that appears to provide a fundamental limit distinguishing quantum behavior from classical behavior. Because of the unique association between critical length and mass, it appears that we can correlate the mass of an object with the size over which its quantum behavior is manifested. When the expression for the critical length is set equal to the Planck length, we find an associated mass value that in magnitude corresponds to an approximation of the mass of the visible universe. This would appear to suggest that the quantum behavior associated with the universe as a whole would be manifested at distances comparable to or smaller than the Planck length. Accordingly, it would appear that all position measurements would be subject to uncertainties at the limit of the Planck length, so that the Planck length sets a fundamental limit on position determination.*

## 1. INTRODUCTION

The Planck length is a natural unit of length defined from three fundamental constants. It is given by:

$$L_P = (hG/2\pi c^3)^{1/2} \qquad (1)$$

where h is Planck's constant, G is the gravitational constant, and c is the speed of light.[1-3] The Planck length has a magnitude of approximately $1.6 \times 10^{-35}$ meter. The Planck length is of particular interest because it combines constants associated with both gravitation and quantum theory, and therefore may be presumed to have a role in quantum gravity.

The Planck length was originally introduced not on the basis of theory but rather on an ad hoc basis from dimensional analysis. Other heuristic approaches to arriving at the Planck length have been used. The significance of the Planck length has been interpreted in terms of the fact that it is roughly equal to the radius of a black hole having a Schwarzschild radius equal to its Compton wavelength.[1] Arguments have been developed on the basis of several approaches including gedanken experiments and dimensional analysis that the Planck length must set a limit on the precision with which an object's position can be measured.[1,4] It is widely considered that at distances of the order of or below the Planck length, fluctuations in the geometry of space-time will be noticeable and dominate the geometry of space-time.[1-4]

We wish to propose a new and different basis for the interpretation of the Planck length, which also leads to the conclusion that quantum fluctuations will be important at length scales below the Planck length, and that the Planck length will set a fundamental limit on position determination.

## 2. CRITICAL LENGTHS AND QUANTUM UNCERTAINTY

Recent work suggests that a possible limiting factor responsible for quantum uncertainty is the presence of the intrinsic spread of Hubble velocities within any extended object. A critical length distinguishing quantum behavior from classical behavior of objects on the basis of this effect has been derived and is given by the equation:[5,6]

$$L_{cr} = [h/4\pi m H_o]^{1/2} \qquad (2)$$

Here, m is the mass of the object, and $H_o$ is the Hubble constant.

An object whose size is smaller than its critical length will exhibit quantum behavior as an entire object, except to the extent that other environmental sources of quantum decoherence may cause it to behave classically. An object whose size is larger than its critical length will exhibit classical behavior, but will also exhibit quantum uncertainties at the level of its critical length. That the quantum uncertainty in the behavior of an object is limited to a region smaller than its critical length seems also to be confirmed by further calculations based on stochastic quantum mechanics.[7,8]

## 3. THE PLANCK LENGTH AS A CRITICAL LENGTH

As part of an exploration of limiting cases, we will now examine what the implications may be of interpreting the Planck length as a critical length.

We can set the Planck length equal to a critical length by combining Eqn. (1) with Eqn. (2). From this, we can evaluate the magnitude of the mass associated with this particular critical length. An expression for the mass associated with a critical length equal to the Planck length is thus given by:

$$m = c^3/(2GH_o) \qquad (3)$$

So what is the significance of this particular mass value? From the magnitudes of the quantities involved, this mass value clearly represents an extremely large mass, something comparable in magnitude to the mass of the visible universe.

We may inquire whether the expression for a mass in Eqn. (3) enters physics in any other context, and whether there is an explicit reason to consider this mass value to be in fact an estimate of the mass of the visible universe.

## 3. ESTIMATION OF THE MASS OF THE VISIBLE UNIVERSE FROM COSMOLOGICAL CRITERIA

There is a certain critical mass density that determines the overall structure of the universe. If the density of the universe is lower than this value, the universe will be infinite (open universe), whereas if the density is greater than this value, the universe will necessarily be finite (closed universe). General relativity predicts that an infinite universe will continue to expand forever, whereas a finite universe will expand for a finite time and then contract. If the mass density has exactly the critical value than the universe is also infinite, and is referred to as flat rather than as open in this case. Contemporary observational evidence indicates that we live in a universe that is expanding and that is flat. Thus, we appear to live in a critical density universe, - in a universe with an average matter density very close to the critical density. Accordingly, the parameter $\Omega$, which represents the fraction of the critical density which is actually present in the universe, seems to be equal to 1 as closely as has been determined.

An expression for the critical density can be calculated in a fairly straightforward manner; it is given by the equation:[9]

$$\rho = 3H_o^2/(8\pi G) \qquad (4)$$

We will use this equation together with a value for the volume of the visible universe in order to estimate the mass of the visible universe. The radius of the visible universe can be estimated as the product of the Hubble time (the inverse of the Hubble constant) with the speed of light; the volume of the visible universe would then be given by the volume of a sphere of that radius. If we take the product of this volume with the critical mass density in the preceding equation, we can obtain an estimate of the mass of the visible universe as:

$$M \approx (4\pi R^3/3)\rho \approx (c^3/2GH_o) \qquad (5)$$

Thus, using these assumptions, we have arrived at an expression for the mass of the visible universe in terms of the gravitational constant, the speed of light, and the Hubble constant.

## 5. MASS ASSOCIATED WITH THE PLANCK LENGTH

Comparing Eqn. (3) with Eqn. (5), we see that the value of the mass associated with the Planck constant as a critical length as given in Eqn. (3), does correspond to the estimate of the mass of the visible universe given in Eqn. (5). Thus, the Planck length appears to be a measure of the critical length associated with the mass of the visible universe.

## 6. INTERPRETING THE PLANCK LENGTH: DISCUSSION AND CONCLUSIONS

We note that the original investigations that introduced the concept of a critical length and addressed its significance were specifically directed toward investigating effects caused by the presence of extremely small recessional velocities throughout extended objects, and the results, including Eqn. (2), were derived from basic principles of quantum theory in that context.[5-8] In the present paper, we started straightforwardly enough, looking beyond the realm of familiar objects in order to evaluate what the equations may be telling us in the case of an extraordinarily small critical length, the Planck length, and we found a prediction that the mass associated with the Planck length would correspond to the mass of the entire visible universe. However, although our approach was straightforward, we implicitly moved beyond the range of validity of the original assumptions on which Eqn. (2) is based. While the results that we have arrived at amount to extrapolations beyond the limitations of the original derivation, they appear to be sufficiently interesting to provide guidance for further analysis.

Earlier results indicated that the quantum behavior of an object is limited to and manifested within a region of size characterized by the critical length.[5-8] This new result therefore seems to be telling us that the quantum behavior associated with the mass of the visible universe would be manifested within distances smaller than the Planck length. Accordingly, it appears that there would be a positional uncertainty associated with the mass of the universe that is approximately equal to the Planck length. Measurement of the position of any object relative to the rest of the universe would therefore be limited in precision by the quantum uncertainty in position associated with the universe as a whole, which is expressed by the critical length determined by the mass of the universe, and hence by the Planck length.

Therefore, we may infer, within the limits of validity of these calculations, that all position measurements will come up against quantum uncertainties at the limit of the Planck length. This analysis thus supports the idea that the Planck length sets a fundamental limit on position determination.

**C. L. Herzenberg**
1700 E. 56th Street #2707
Chicago, IL 60637-5092 U.S.A.
e-mail: carol@herzenberg.net

interpretation planck length.doc
16 October 2006 draft, revised